\documentclass[aps,pra,twocolumn,footinbib,
showpacs,showkeys,nobalancelastpage]{revtex4-2}

\usepackage{graphicx}
\usepackage{indentfirst}
\usepackage{physics}
\usepackage{braket}
\usepackage{float}
\usepackage{amsmath}
\usepackage{amssymb}
\usepackage{verbatim}
\usepackage{epstopdf}
\usepackage{CJK}
\usepackage{esint}
\usepackage{color}
\usepackage{epsfig}
\usepackage{subfigure}
\usepackage{soul}
\usepackage{amsfonts}
\usepackage{footmisc}
\usepackage{scrextend}
\usepackage{multirow}
\usepackage{mathtools}
\usepackage{xr-hyper}
\usepackage[hyperfootnotes=false]{hyperref}

\usepackage[english]{babel}
\usepackage{url}
\usepackage{bm}
\definecolor{darkblue}{rgb}{0,0,0.5}
\hypersetup{
colorlinks=true,
linkcolor=black,
filecolor=black,
citecolor=darkblue,
urlcolor=black,
}

\newcommand{\defeq}{\vcentcolon=}

\newcommand\bs[1]{\boldsymbol{#1}}
\newcommand\mc[1]{\mathcal{#1}}
\newcommand\kCPF[1]{\mathcal{U}_{\text{TPF}}^{#1}}

\begin{document}

\title{Secure Quantum Pattern Communication}
\author{Cillian Harney}
\email{cth528@york.ac.uk}
\author{Stefano Pirandola}
\email{stefano.pirandola@york.ac.uk}
\affiliation{Department of Computer Science, University of York, York YO10 5GH, UK}
\begin{abstract}
We propose a multi-mode modulation scheme for Continuous Variable (CV) quantum communications, which we call quantum pattern encoding. In this setting, classical information can be encoded into multi-mode patterns of discretely-modulated coherent states, which form instances of a communicable image space. Communicators can devise arbitrarily complex encoding schemes which are degenerate and highly non-uniform, such that communication is likened to the task of pattern recognition. We explore initial communication schemes that exploit these techniques, and which lead to an increased encoding complexity. We discuss the impact that this has on the role of a near-term quantum eavesdropper; formulating new, realistic classes of attacks and secure communication rates.
\end{abstract}
\maketitle

The rapid maturation of the field of quantum communications \cite{AdvCrypt,UniteQInt} promises to make it one of the first technologies to be featured in the upcoming quantum revolution. By exploiting quantum information theoretic protocols \cite{Mike_Ike, WatrousTxt, Holevo19}, we can assure provably secure communication based on underlying physical principles. Protocols that utilise Continuous Variable (CV) quantum systems \cite{SerafiniCV,GaussRev,BraunsteinVL} (such as bosonic modes) form a particularly promising area of research \cite{RalphCV}, thanks to their high performance, near-term practical feasibility, and potential for large scale deployment using current telecommunication infrastructures. \par
There exist a wide variety of protocols derived from CV encodings, many of which rely on a continuous-modulation (Gaussian) of various Gaussian states \cite{Coherent1,Coherent2,Squeezed1,Squeezed2,Thermal1,Thermal2}. Over the years, rigorous security proofs have been obtained for these protocols, alongside theoretical/experimental evidence of their efficacy \cite{PLOB,CVQKDGen,CVQKD202}. However, the study of discretely-modulated CV systems is also of significant interest, where finite-dimensional entities are embedded into infinite-dimensional Hilbert spaces \cite{PulsedHomodyne,Grangier1,SychMPSK,PanosPE1,Grangier2,PanosPE2}. Such discrete-modulation schemes present simplifications over Gaussian-modulation with regards to state preparation and data processing. \par
Alternative modulation schemes can be devised when considering multiple bosonic modes. The use of multi-mode technologies has been shown to be advantageous in a number of quantum communication settings \cite{FLQKD1,FLQKD3,EnhancedECQC}, where communicable symbols are encoded into multiple modes, or via repeated channel usage. In Refs.~\cite{PPM1,PPM2,OptPGM} the authors study the utility of highly symmetric collections of multi-mode, binary-modulated coherent states, whose optimal discrimination is easier to obtain globally rather than locally. In this way, highly efficient communication schemes can be based on the packaging of $d$-ary variables into multi-mode coherent states. \par
Yet, multi-mode encoding invites a further abstraction. Let us define a \textit{quantum pattern} as a $m$-mode coherent state undergoing local, $k$-ary modulations. It is possible to construct a collection of quantum patterns that belong to a global \textit{image space}, forming a sub-set of all $k^m$ possible patterns that exist. Each element of this image space can be endowed with a particular \text{classification} that encodes a communicable symbol; embedding information
not into local modulations, but into an abstract classification process associated with pattern features.\par
This marks a significant departure from any form of encoding used in standard communications. If information can be encoded into conceptual properties of a coherent pattern space, it is possible to impose extreme classification degeneracies and non-linearities; aligning the tasks of communication and pattern recognition very closely. Codes can be designed that exploit specific multi-mode technologies, or embed extractable features into vast data-sets. 
Furthermore, the recent integration of modern machine learning tools within quantum hypothesis testing \cite{nielsenNN,probML,PatternRecog, ThermalPatt} further encourages an application of these methods to quantum communication.\par
The introduction of quantum pattern encoding also raises interesting questions about realistic eavesdroppers and security. While unconditional security must consider an eavesdropper with unlimited resources, perfect quantum memories and a full working knowledge of the protocol, these assumptions may not realistically hold in the presence of overwhelmingly complex (possibly data-driven) codes. Hence the application of versatile, machine-learning enhanced encoders/receivers may be used to cast doubt on the knowledge of an attacker, and improve communication rates. This makes it non-trivial to consider scenarios of information asymmetry between trusted-parties and eavesdroppers. In this work, we explore this asymmetry by devising new, weaker classes of eavesdropper attacks which may emerge within the pattern communication regime.\par
This paper proceeds as follows: In Section \ref{sec:Prelims} we explicitly introduce coherent quantum patterns. In Section \ref{sec:SecureSec}, we provide a general approach to studying secure communication, establishing a hierarchy of rates based on eavesdropper resources. In Section \ref{sec:PatternEncs}, we devise two binary pattern encoding schemes and illustrate their performance over pure-loss channels. Finally, we provide concluding discussions and possible future investigative paths.

\section{Quantum Pattern Communication\label{sec:Prelims}}

\subsection{Coherent Quantum Patterns}

Let us formally {introduce the concept of a} \textit{coherent quantum pattern}. {This is a} discrete ensemble of coherent states that undergo $k$-ary modulation. Let $\bs{i} = \{i_1, i_2,\ldots,i_m\}$ denote an $m$-length string where {each element of the string is a random variable that can occupy $k$ unique values}, $i_j \in \{0,\ldots,k-1\}$. This string (or pattern) {can be used to generate} a corresponding coherent pattern state given by
\begin{gather}
\begin{gathered}
\ket{\alpha_{\bs{i}}} \defeq \ket{\alpha_{i_1}} \otimes \ket{\alpha_{i_2}} \otimes \ldots \otimes \ket{\alpha_{i_m}} = \bigotimes_{j=1}^m \ket{\alpha_{i_j}}, \\
\text{where } \ket{\alpha_{i_j}} \in \{\ket{\alpha_0}, \ket{\alpha_1}, \ldots, \ket{\alpha_{k-1}}\}. 
\end{gathered}
\label{eq:CohPatS}
\end{gather}
For example, if $k=2$ {then we are employing a binary modulation on each local mode. In this case, one can utilise BPSK so that each local coherent state $\ket{\alpha_{i_j}}$ will take the form $\ket{\alpha_{i_j}} \in  \{\ket{\sqrt{N_S}}, \ket{-\sqrt{N_S}}\}$, or Binary Amplitude Modulation (BAM) where $\ket{\alpha_{i_j}} \in \{\ket{0}, \ket{\sqrt{N_S}}\}$, where $N_S$ denotes the mean number of photons transmitted in each state \cite{GaussRev}.} 

{A coherent pattern state $\ket{\alpha_{\bs{i}}}$ represents a single state generated by the pattern $\bs{i}$. However, our goal is to create a basis for quantum communication, and we therefore require much more than just a single $\bs{i}$. To this end, we define \textit{an image space} as a collection of many $k$-ary patterns which are used to generate a potentially vast collection of coherent pattern states. More precisely, an $N$-element image space can be used to generate a corresponding collection of coherent pattern states},
\begin{equation}
\mc{U} \defeq \{ \bs{i}_1, \bs{i}_2, \ldots, \bs{i}_N \} \rightarrow \{  \ket{\alpha}_{\bs{i}} \}_{\bs{i}\in\mc{U}}.
\end{equation}
{This collection of states can then be used to create a basis for quantum communications. For $m$-mode patterns undergoing $k$-ary local modulations, the set of all possible patterns contains $k^m$ elements. }

{Crucially, coherent patterns can be used to formulate a mapping between a $d$-dimensional alphabet $\mathcal{A} = \{1,\ldots,d\}$ which contains symbols used to construct secret-keys}, and an image space $\mathcal{U}$. Each pattern $\bs{i} \in \mc{U}$ can be used to represent a symbol from the alphabet, which is the same as assigning a specific classification $c\in \mathcal{A}$ to each pattern. {The formal mapping between an image space and an alphabet} is described by a codebook $\mathcal{C}$, which formally takes the form
\begin{equation}
\mathcal{C} \defeq \left\{ \big(c(\bs{i})~ ; \ket{\alpha_{\bs{i}}}\big) ~\big|~ c(\bs{i}) \in \mathcal{A}, \bs{i} \in \mathcal{U} \right\}, \label{eq:GenCodebook}
\end{equation}
where $c(\bs{i}) = c \in \mathcal{A}$ is the classification of a pattern $\bs{i}$. {The alphabet and codebook thus completely characterises the pattern modulation scheme; a sender (Alice) may transmit pattern states to a receiver (Bob) who must then discriminate the incoming pattern and its classification can be inferred by consulting with the shared codebook. We may refer to a pattern encoding setup using the alphabet and codebook tuple $(\mc{A},\mc{C})$.}

The construction of an image space $\mathcal{U}$ is incredibly flexible. It is by no means compulsory that the alphabet and image space are of the same dimension $|\mathcal{A}| = |\mathcal{U}|$, i.e.~the encoding need not be a one-to-one mapping between patterns $\bs{i}\in \mathcal{U}$ and symbols $c\in \mathcal{A}$. In general, each symbol maps to a subset of the image space, $c \mapsto \{\bs{i}\in \mathcal{B}(c)\} \subset \mathcal{U}$, meaning that an image space can be decomposed according to class equivalent subsets,
\begin{equation}
\mathcal{U} = \bigcup_{c\in\mathcal{A}} \mathcal{B}(c),~~ \mathcal{B}(c) = \{\bs{i} \in \mathcal{U}~|~c(\bs{i}) = c\} \label{eq:ClassPatts}.
\end{equation}
Each subset $\mathcal{B}(c)$ is filled with many potential codewords of varying forms, however they should all possess abstract features that allow them to be classified as belonging to the class $c$.
Furthermore, these subsets do not necessarily have a well defined size, but in reality we must possess a finite set of samples from which classifiers can draw expertise. For a $d$-dimensional alphabet, if each class of pattern state is transmitted with equal \textit{a priori} probability $p_c = 1/d$, then the probability of transmitting any single pattern is, $p_{c(\bs{i})}= 1/({d |\mathcal{B}({c(\bs{i})})|})$. An encoding of this form is called degenerate, and is explored in subsequent sections. 

\subsection{Practical Aspects}
We may consider two realisations of quantum pattern {transmissions}, corresponding to one dimensional (1D) and two dimensional (2D) patterns. This depends entirely on the spatio-temporal configuration that one is interested in. A 1D pattern transmission corresponds to single-shot multi-mode communications: Alice encodes information into quantum states simultaneously transmitted over $m$-spatial modes to Bob.
Hence, any symbol is transmitted via the single use of each spatial mode. However, this may be quite restrictive, as the use of large pattern encodings leads to a potentially unfeasible number of spatial modes.\par
Instead, Alice may perform 2D pattern transmissions, achieved by introducing temporal extensions of each spatial mode; corresponding to multi-shot communications. Alice and Bob now communicate over a fixed time period $T$ seconds, discretising this period into time-bins. Alice transmits information sequentially over $m^\prime$-spatial modes and $m^{\prime \prime}$-temporal modes, such that each transmission point in space and time corresponds to a $k$-ary variable. This allows Alice to construct an $m = (m^{\prime}\times m^{\prime \prime})$ length pattern. \par
Importantly, any 2D pattern using identical $m^{\prime}$-spatial modes, and $m^{\prime \prime}$-temporal modes can always be expanded to a 1D pattern with exclusively $m$-spatial modes, and vice versa. This is provided through an assumption of uniform channels; if the multi-channel is not uniform, then degenerate channels can be pooled together, or expanded in a similar way.

\subsection{Pattern Modulated CV-QKD, Communication Rates and Security \label{sec:SecureSec}}

{Standard discretely-modulated CV-QKD uses phase/amplitude encoded coherent states which form constellation. Alice randomly generates and transmits coherent states from this constellation to Bob, followed by parameter estimation, privacy amplification, error-correction etc.~in order to establish a secret key. In this protocol the mapping between each discretely-modulated coherent state and its binary presentation is \textit{public}. This means that if an attacker (Eve) intercepts a state and correctly discriminates it as a particular coherent state from the constellation, she can correctly extract its binary representation use for key generation. }

{Quantum pattern communication is a generalised modulation scheme where we possess a mapping between an abstract image space of coherent pattern states $\{\alpha_{\bs{i}}\}_{\bs{i}\in\mc{U}}$ and a set of corresponding symbols, contained in the codebook $\mc{C}$. The relationship between a quantum pattern and its symbol in the codebook may be highly non-linear and degenerate as it is encoded into global, multi-modal features. This produces a communication basis which is used for key distribution, i.e.~Alice randomly generates quantum patterns from the image space which are transmitted and discriminated by Bob, followed by the standard CV-QKD steps in order to establish a secret-key. }

Like many other discretely modulated bosonic communication schemes, evaluating the efficacy (secure communication rate) of pattern-based protocols can be very demanding. In a best case scenario we would study information transmission over thermal-loss channels. However the addition of thermal noise to the already non-Gaussian ensemble of discretely modulated coherent states requires treatment in an infinite dimensional, multi-mode Fock space that is computationally infeasible. For this reason, we restrict our studies to bosonic pure-loss channels $\mathcal{E}_{\eta}$ with transmissivity $\eta$ as an initial step in the study of this topic. For the transmission of $m$-mode pattern states we assume that these channels are uniform, such that $\mathcal{E}_{\eta}^m \defeq \bigotimes_{j=1}^m \mathcal{E}_{\eta}$ \footnote{This assumption is strong when considering a small number of spatial modes}. Here we study secure quantum communication rates {assuming the use of CV-QKD} in direct reconciliation. Thus we focus on a one-way, sender (Alice) to receiver (Bob) scheme, subject to an eavesdropper (Eve). \par

Consider the use of an encoding scheme $(\mathcal{A},\mathcal{C})$ using Eq.~(\ref{eq:GenCodebook}) over uniform pure-loss channels. In the asymptotic regime of many exchanged signals, we may quantify the performance of a communication protocol through the following secure transmission rate \cite{CoverThomas,Csiszar},
\begin{gather}
R \defeq I_{AB} - I_{AE} ,
\end{gather}
where $I_{AB}$ is the mutual information between the parties Alice and Bob, while $I_{AE}$ measures Eve's ability to extract information from the protocol (we also assume ideal reconciliation of data for Alice and Bob). The form of $I_{XY}$ for the respective parties depends on multiple factors; in particular, Eve's performance depends directly on the level of threat that she poses. \par

The maximum amount of classical information accessible to Bob or Eve is upper bounded by their Holevo information. Assuming all symbols are transmitted with equal \textit{a priori} probability $p_{c(\bs{i})}$, and defining $\alpha_{\bs{i}}^{\eta} \defeq \ketbra{\eta \alpha_{\bs{i}}}$, we may write
\begin{gather}
I_{AB} \leq \chi_{AB}(\eta) , \>\> I_{AE} \leq \chi_{AE}(1-\eta) ,\\
\chi(\eta) \defeq S\left(\sum_{\bs{i}\in\mathcal{U}} p_{c(\bs{i})} \alpha_{\bs{i}}^{\eta} \right) - \sum_{\bs{i}\in\mathcal{U}} p_{c(\bs{i})} S( \alpha_{\bs{i}}^{\eta}),
\end{gather}
where $S(\cdot)$ denotes the von Neumann entropy. Eve's maximum potential threat is safely quantified by the Holevo bound $\chi_{AE}$, making communication unconditionally secure provided $I_{AB} \geq \chi_{AE}$. This bound assumes that Eve applies a beam-splitter attack, followed by perfect storage of the stolen modes in a quantum memory prior to information extraction (an entangling-cloner attack). While this is compulsory to ensure unconditional security, it is not always a realistic assumption on behalf of near-term quantum eavesdroppers.\par

Pattern encoding introduces a novel twist on traditional security assumptions. Typical quantum communications scenarios (discretely or continuously modulated) embed classical information in such a way that the mapping between quantum states and their classical symbols are either public, or reliably inferred by an interceptor. For instance, if Eve intercepts the communications of a four-state, phase-encoded protocol, even if \textit{a priori} unknown the encoding can be easily inferred over a number of transmissions.

If one utilises pattern encoding, this inference is no longer a trivial assumption, but an additional obstacle for Eve to overcome. {Alice and Bob either (\textit{i}) engage in a pre-communication secure training protocol in order to construct an effective classifier for an encoding scheme \cite{MultilabelCVQKD,QuantumSecureLearning}, or (\textit{ii}) share a pre-agreed, precise codebook to be used for communication. The increased complexity of the encoding scheme means that in a practical setting, it is extremely unlikely that Eve will have, or deduce, perfect knowledge of the codebook.} This invites a new class of weak but realistic attacks on a pattern communication protocol, which we label \textit{approximate attacks}. These attacks emerge from asymmetry between the pre-determined encoding chosen by Alice and Bob $(\mathcal{A}, \mathcal{C})$, and that which Eve has access to, $(\mathcal{A}_{E}, \mathcal{C}_{E})$. In this way we may establish a new hierarchy of eavesdropper threats, from approximate to collective attacks.\par

\subsection{Mutual Information}

Consider a pattern encoding $(\mathcal{A}, \mathcal{C})$, where $\mathcal{C}$ is constructed from some appropriate image space $\mathcal{U}$. Alice transmits a pattern $\bs{i}_A$ with classification $c(\bs{i}_A)$. Bob has full knowledge of the encoding scheme, and can therefore optimise his (generally quantum) measurements such that any incoming, noisy pattern $\bs{i}_B$ is classified according to a set of optimised POVMs $\tilde{\bs{\Pi}} \defeq \big\{\tilde{\Pi}_{{c}(\bs{i})}\big\}_{{c}({\bs{i})\in\mathcal{A}}}$. Measurements of this form $\tilde{\bs{\Pi}}$ are designed in such a way that discrimination of the pattern $\bs{i} \in \mathcal{U}$ and classification ${c}(\bs{i}) \in \mathcal{A}$ are combined in a cohesive process, and may be achieved via fully coherent, quantum algorithms. That is, an input pattern would be processed by an optimised quantum circuit followed by some projective measurement onto the assigned class. For highly complex/non-linear encodings, this task is best addressed by quantum machine learning \cite{QMLGen,PreskillQML}. \par
Yet, in the absence of fully coherent class measurements, this task can be more simply split into quantum pattern discrimination $\bs{\Pi} \defeq \left\{\Pi_{\bs{i}}\right\}_{\bs{i}\in\mathcal{U}}$ followed by classical post-processing via a classifier $\tilde{c}$, such that 
$\tilde{c}(\bs{i}) \in \mathcal{A}$ {denotes the class prediction of a pattern $\bs{i}$ according to this classifier.} Indeed, this classification process aligns itself with near-term, realistic resources, providing access to powerful, modern pattern recognition tools. The goal of communication is to maximise the probability {that the classifier's prediction of the received pattern is equal to the class of the initial pattern}, i.e.~$\tilde{c}(\bs{i}_B) \approx c(\bs{i}_A)$. Imposing a choice of classifier ${\tilde{c}}$, the conditional probabilities take the form,
\begin{align}
p_{\tilde{c}}(c_B| c_A) &= \sum_{\bs{i}_A \in \mathcal{B}(c_A),\bs{i}_B \in \mathcal{B}(c_B)}  \hspace{-3mm} p_{\tilde{c}}(c_B | \bs{i}_B) \text{Tr}\left[ \Pi_{\bs{i}_B} \alpha_{\bs{i}_A}^{\eta} \right], \\
p_{\tilde{c}}(c_A| c_B) &= \frac{ p_{\tilde{c}}(c_A, c_B)}{ p_{\tilde{c}}(c_B)} =  \frac{p_{\tilde{c}}(c_B| c_A)}{\sum_{c_A\in \mathcal{A}} p_{\tilde{c}}(c_B| c_A)},
\end{align}
where the second line follows from Bayes theorem. \par
Assuming {a pattern class is transmitted with equal probability of any other class}, the mutual information between Alice and Bob then takes the form, 
\begin{align}
 I_{AB}^{\bs{\Pi},\tilde{c}}(\eta)= \log(|\mathcal{A}|) + \hspace{-4mm} \sum_{c_A, c_B\in \mathcal{A}}\hspace{-2.5mm} p_{\tilde{c}}(c_A, c_B) \log(p_{\tilde{c}}({c_A|c_B})). \label{eq:MutInfoAB}
\end{align}
Throughout this work $\log$ is taken as base 2. This quantifies Alice and Bob's information retrieval given Bob's perfect knowledge of the encoding, and the split measurement-classification process using the POVM set $\bs{\Pi}$ and statistical classifier $\tilde{c}$. It also provides an alternative way to upper bound the mutual information in the absence of coherent class measurements, 
\begin{equation}
I_{AB} \leq \max_{\bs{\Pi}, \tilde{c}} \left( I_{AB}^{\bs{\Pi},\tilde{c}} \right) \leq \max_{\tilde{\bs{\Pi}}} \left( I_{AB}^{\tilde{\bs{\Pi}}} \right) \leq \chi_{AB}.
\end{equation}
When a one-to-one encoding is used, pattern classifications and the patterns themselves are equivalent and therefore, Eq.~(\ref{eq:MutInfoAB}) simplifies without the need for a classifier.

\subsection{Security Hierarchy}
We are now in a position to develop a security hierarchy. Consider communication such that Alice and Bob utilise an encoding $(\mathcal{A}, \mathcal{C})$, achieving the realistic transmission rate in Eq.~(\ref{eq:MutInfoAB}). We now introduce an eavesdropper with a (potentially different) encoding $(\mathcal{A}_E, \mathcal{C}_{E})$. {Enhanced security} hinges on the asymmetry between these, and we will discuss this hierarchy in order of decreasing threat.\par
As discussed, unconditional security is guaranteed through the assumption of Eve's access to quantum memories, and perfect knowledge of the encoding such that $(\mathcal{A}_E, \mathcal{C}_{E}) = (\mathcal{A}, \mathcal{C})$. In this general setting of collective attacks and perfect knowledge, the rate can be lower bounded according to
\begin{equation}
R_{\text{coll}} = I_{AB}^{\bs{\Pi},\tilde{c}}(\eta)- \chi_{AE}(1-\eta). \label{eq:GenRate}
\end{equation}
Hence under collective attacks, communication is only secured via high transmissivity, $\eta > 0.5$ \cite{AdvCrypt}. A more realistic rate for near term technologies (but less secure) is achieved by removing Eve's ability to extract the accessible information. Granting Bob and Eve identical measurement apparatus and classifiers $(\bs{\Pi}_B,\tilde{c}_B) = (\bs{\Pi}_E,\tilde{c}_E) = (\bs{\Pi},\tilde{c})$, then we may consider a rate proposed by individual attacks,
\begin{equation}
R_{\text{ind}} = I_{AB}^{\bs{\Pi},\tilde{c}}(\eta) - I_{AE}^{\bs{\Pi},\tilde{c}}(1-\eta) \geq R_{\text{coll}}. \label{eq:IndRate}
\end{equation}
Bob and Eve's performances are symmetric with respect to transmissivity, therefore secure communication is limited to $\eta \geq 0.5$ \cite{AdvCrypt}.\par

We may consider weaker classes of attacks by removing this symmetry. There exist scenarios where Eve will not possess perfect knowledge of the encoding scheme, $(\mathcal{A}_E, \mathcal{C}_{E}) \neq (\mathcal{A}, \mathcal{C})$ due to the complexity of the pattern communication regime. This can be hugely detrimental to Eve, as even minute inaccuracies in her codebook or alphabet can have a significant impact on her information retrieval. Generally, Eve's ignorance to the correct encoding leads to a codebook of the form,
\begin{equation}
\mathcal{C}_E = \left\{ \big(c(\bs{i})~ ; \ket{\alpha_{\bs{i}}}\big) ~\big|~ c(\bs{i}) \in \mathcal{A}_E, \bs{i} \in \mathcal{U}_E \right\}, \label{eq:EveCodebook}
\end{equation}
where $\mathcal{U}_E \neq \mathcal{U}$ is a sub-optimal image space of \textit{potential} pattern states, and may be larger or smaller than $\mathcal{U}$ dependent on the scenario. We define an \textit{approximate attack} as an individual attack by an eavesdropper who possesses only partial knowledge of the encoding. We denote approximate attack rates using $\tilde{R}$, and once more progress in order of decreasing threat.\par
Consider a degenerate encoding scheme $(\mathcal{A},\mathcal{C})$, and an approximate attack in which Eve is aware of the alphabet to codebook mapping, but possesses a sub-optimal image space of potential patterns. That is, $\mathcal{A}_E = \mathcal{A}$, but for the image space
\begin{equation}
 \mathcal{U}_E = \bigcup_{c\in\mathcal{A}} \mathcal{B}_{E}(c) \text{ , } \exists~c \in \mathcal{A} \text{ s.t } |\mathcal{B}_{E}(c)| < |\mathcal{B}({c})|, 
\end{equation}
where $\mathcal{B}(c)$ are subspaces of class equivalent patterns as in Eq.~(\ref{eq:ClassPatts}). That is, Eve is missing potential elements of the degenerate image space. In the limit of maximum ignorance, Eve possesses only one example of each class codeword $|\mathcal{B}_{E}(c)| = 1$, $\forall c\in\mathcal{A}$. Since Eve is still knowledgeable of the encoding, she may optimise her measurement apparatus ($\bs{\Pi}_E = \bs{\Pi}$). But the diminished image space renders her classifier $\tilde{c}_E$ inferior with respect to Bob's, since there is less expertise to draw from the reduced image space $\mathcal{U}_E$. More formally, Eve's expected error rate of classification over a set of pattern transmissions $\bs{i} \in \mathcal{V}$ may be substantially worse than Bob's,
\begin{equation}
\mathbb{E}_{\mathcal{V}} \left[ p(c(\bs{i})|\bs{i},\mathcal{U}_E)\right] < \mathbb{E}_{\mathcal{V}} \left[ p(c(\bs{i})|\bs{i},\mathcal{U})\right].
\end{equation}
We label this as a \textit{diminished approximate attack}, leading to the new rate,
\begin{equation}
\tilde{R}_{\text{dim}} = I_{AB}^{\bs{\Pi},\tilde{c}}(\eta) - I_{AE}^{\bs{\Pi},\tilde{c}_E}(1-\eta)  \geq R_{\text{ind}}. \label{eq:DimRate}
\end{equation}
Summarising, these are a form of individual attack in which Eve's resources limit her ability to optimise a classifier. For one-to-one pattern encodings, there exist only one example of each class codeword anyway, hence this attack is no longer approximate and $\tilde{R}_{\text{dim}} = R_{\text{ind}}$.\par
The previous attack assumed that Eve still retained knowledge of the codeword to symbol mapping, however, for larger code-spaces and alphabets it is possible to construct pattern embeddings which are close to indistinguishable from other codes. This makes codebook/alphabet inference extremely difficult. Consider an approximate attack such that Eve is in possession of sub-optimal image space which is larger than Alice and Bob's $\mc{U} \subset \mathcal{U}_E$, and must use this to infer the correct encoding to retrieve any information. Since $\mathcal{U}_E$ is larger than $\mathcal{U}$ it contains potentially invalid patterns, meaning $\bs{\Pi}_E$ and $\tilde{c}_E$ will also become sub-optimal. Furthermore, the attack is now probabilistic, since there is a chance that she will infer an incorrect encoding. Given that Eve can successfully learn $(\mathcal{A}, \mathcal{C})$ with some probability $p_{\text{dec}}$, we obtain the rate
\begin{equation}
\tilde{R}_{\text{pr}} = I_{AB}^{\bs{\Pi},\tilde{c}}(\eta) - p_{\text{dec}} \left(I_{AE}^{\bs{\Pi}_E,\tilde{c}_E}(1-\eta)\right)  \geq R_{\text{ind}}.
\end{equation}
This is a \textit{probabilistic approximate attack}, and describes a situation in which codeword to alphabet mappings cannot be trivially obtained by an eavesdropper. For large, multi-mode encodings, $p_{\text{dec}}$ can be made extremely small dependent on how much encoding information has been leaked to Eve. This formulates our weakest class of attack for pattern communications, allowing for the hierarchy,
\begin{equation}
{R}_{\text{coll}} \leq R_{\text{ind}} \leq \tilde{R}_{\text{dim}} \leq \tilde{R}_{\text{pr}} .\label{eq:HierRates}
\end{equation}

\section{Pattern Encoding Schemes \label{sec:PatternEncs}}
In this section we offer a pair of simple introductory examples of {binary-pattern modulated quantum communications, setting $k=2$ and utilising BPSK to construct our coherent pattern bases. That is, we construct $m$-mode coherent quantum patterns $\ket{\alpha_{\bs{i}}} = \bigotimes_{j=1}^m \ket{\alpha_{i_j}}$ using a local binary modulation on each mode, such that each local coherent state is attributed to a \textit{background state} so that $i_j = 0$ and $\ket{\alpha_0} = \ket{ -\sqrt{N_S}}$, or a \textit{target state} so that $i_j = 1$ and $\ket{\alpha_1} = \ket{\sqrt{N_S}}$.}
 We illustrate how the abstraction to global encoding can severely impact the threat of a near term eavesdropper, studying the hierarchy of rates depicted in Eq.~(\ref{eq:HierRates}).

\subsection{Localised-TPF Pattern Modulation \label{sec:LTPF}}
It is known that the discrimination of ensembles of quantum states with Geometrical Uniform Symmetry (GUS) can be enhanced through the use of joint quantum measurements \cite{EldarPGM}. An ensemble of quantum states $\{p_{i} ; \rho_{i}\}_{i=1}^{n}$ {(a collection of states $\rho_{i}$ each of which occur with probability $p_i$)} possess GUS if $p_i = 1/n$ and there exist a set of symmetry unitaries $\{S_i\}_{i=1}^{n}$ which can transform each state $\rho_i$ into another state from the ensemble, $\rho_i = S_i \rho_0 S_i^{\dag}$, and $S_0 = I$ where $I$ is the identity. 
In the case of GUS ensembles of pure coherent states, Pretty Good Measurements (PGMs) have been proven to be optimal {discriminatory measurements} \cite{OptPGM}. This means that GUS ensembles of coherent states transmitted through pure-loss channels (which retain the purity of input states $\mathcal{E}_{\eta}(\ket{\alpha}\!\bra{\alpha}) =\ket{\eta \alpha}\!\bra{\eta \alpha}$) can be optimally discriminated via PGMs.

Motivated by this fact, and inspired by the Channel Position Finding (CPF) formalism developed for quantum channel discrimination \cite{EntEnhanced}, here we introduce the concept of $k$-Target Position Finding ($k$-TPF). {This is an encoding scheme based on the use} of image spaces $\kCPF{m,k}$ which describe the set of all $m$-length binary patterns that possess exactly $k$-target modulated states. For example, if $k=1$ then the image space $\kCPF{m,1}$ denotes the ensemble of $m$-mode coherent states with a single target state, against a backdrop of $(m-1)$ background states. For an explicitly example, take $m=3$ and we could construct the image spaces, 
\begin{align}
\kCPF{3,1} &\defeq \{ \{1,0,0\}, \{0,1,0\}, \{0,0,1\} \}, \\
\kCPF{3,2} &\defeq \{ \{1,1,0\}, \{1,0,1\}, \{0,1,1\} \}.
\end{align}
This form of image space can be be used to generate GUS coherent pattern ensembles for communication between Alice and Bob, as explored in \cite{PPM1}.

\subsubsection{Pattern Modulation Scheme}
We may now outline a potential pattern modulation scheme over uniform, $m$-length multi-channels. Alice and Bob wish to globally encode information onto their $m$-mode patterns by means of two characteristics; locality and TPF properties (number of target modes). 
Any $m$-mode coherent pattern can be divided into an $n$-partite locality structure which identifies particular regions of the pattern state that will have specific characteristics. This partitioning can be described by a disjoint partition set $\mc{S}$ which collects specific modes within the pattern. More precisely, we can construct this disjoint partition set as
\begin{gather}
\mathcal{S} = \{ \bs{s}_1, \bs{s}_2, \ldots, \bs{s}_n \} = \bigcup_{j=1}^{n} \{ \bs{s}_j \},\\
1 \leq |\bs{s}_j| \leq m, \text{ and } \bs{s}_{j} \cap \bs{s}_{k} = \varnothing, \> \forall j\neq k . \label{eq:modes_disjoint}
\end{gather}
Importantly,  $\{1,\ldots,m\} \subseteq \mathcal{S}$ meaning that all $m$-modes are accounted for in the locality structure. Meanwhile, Eq.~(\ref{eq:modes_disjoint}) ensures that only mode labels from 1 to $m$ are considered, and that all sub-collection of modes $\bs{s}_j$ are pairwise disjoint. 

Concurrently, Alice and Bob can assign a $k$-TPF property to each sub-collection of modes. They may construct a $k$-TPF partition set $\mc{K}$ which informs Alice and Bob of how many target modulated states are permitted within any particular sub-region of the quantum pattern state specified by $\mc{S}$. This partition set takes the form,
\begin{gather}
\begin{gathered}
\mathcal{K} = \{ {k}_1, \ldots, k_j, \ldots, k_n \}, \text{ }  {k}_j \in \{1,\ldots,|\bs{s}|_j -1 \}, 
\end{gathered}
\label{eq:modes_disjoint}
\end{gather}
This then ensures that a given sub-pattern $\bs{s}_j$ will contain exactly $k_j$-target modes. Note that $ {k}_j \in \{1,\ldots,|\bs{s}|_j -1 \}$ ensures that at least a binary variable is encoded into each sub-pattern. Finally, Alice and Bob can impose a cardinality condition on their choice of target numbers in each sub-region. Letting where ${C_n^k} = \frac{n!}{k!(n-k)!}$ is the binomial coefficient, then they may impose that
\begin{equation}
{C_{|\bs{s}_1|}^{k_1}} \cdot {C_{|\bs{s}_2|}^{k_2}} \cdot \ldots \cdot {C_{|\bs{s}_n|}^{k_n}}  = \prod_{j=1}^r {C_{|\bs{s}_j|}^{{k}_j}} = \Sigma,
\end{equation}
to ensure that they can communicate exactly $\Sigma$ bits per global transmission.

A global image space can thus be constructed according to
\begin{equation}
\kCPF{\mathcal{S},\mathcal{K}}  = \{ \kCPF{|\bs{s}|_1,k_1}, \ldots, \kCPF{|\bs{s}|_n,k_n} \} = \bigcup_{j=1}^n \kCPF{|\bs{s}|_j,k_j},
\end{equation}
as a concatenation of all the $k_j$-TPF image spaces of each sub-pattern. Hence, we can define a one-to-one encoding in conjunction with these partition sets, with a $\Sigma$-dimensional alphabet $\mathcal{A} = \{1,\ldots,\Sigma\}$, and the following codebook,
\begin{equation}
\mathcal{C} = \left\{ \big( c ~ ; \ket{\alpha_{\bs{i}}}\big) ~\big|~ c \in \mathcal{A}, \bs{i} \in \kCPF{\mathcal{S},\mathcal{K}} \right\}. \label{eq:kTPFCodebook}
\end{equation}

We label this a \textit{Localised Target Position Finding} (LTPF) encoding scheme. Given this information, Bob can \textit{always} optimise his measurement apparatus using optimal POVMs over specific sub-patterns of the global message, in order to discriminate and decode the transmission.  Let us define $\{ \Pi_{\bs{i}}^{m,k} \}_{\bs{i} \in \kCPF{k}}$ as the optimal set of PGMs for discriminating an $m$-mode, $k$-TPF pure state ensemble. Then for an $(\mathcal{S},\mathcal{K})$ encoding scheme, we utilise the following set of optimal POVMs,
\begin{equation}
\bs{\Pi}^{\mathcal{S},\mathcal{K}} = \left\{ \Pi_{\bs{i}} \right\}_{\bs{i}\in\kCPF{\mathcal{S},\mathcal{K}} } , \>\> \Pi_{\bs{i}}^{\mathcal{S},\mathcal{K}} = \bigotimes_{j=1}^{n} \Pi_{\bs{i}^{\bs{s}_j}}^{|\bs{s}_j|,k_j},
\end{equation}
where $\bs{i}^{\bs{s}_j}$ denotes the sub-pattern corresponding to the modes contained in the $j^{\text{th}}$ partition, $\bs{s}_j$. See Fig.~\ref{fig:LTPF_rates}(a) for an example of this communication setup. 

\begin{figure}
\includegraphics[width=\linewidth]{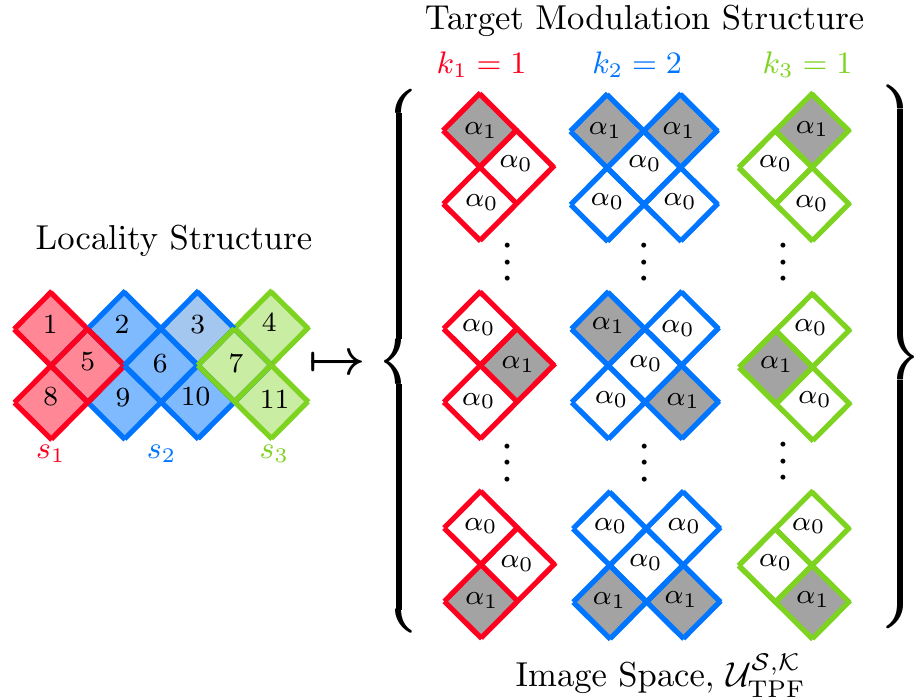}\\
\caption{{The LTPF modulation scheme using $m$-mode coherent quantum patterns. This illustrates an example for $m=11$, where the locality structure is described by a disjoint collection of modes $\mc{S} = \{ \{1,5,8\},\{2,3,6,9,10\},\{4,7,11\}\}$ and an associated $k$-TPF property assigned to each collection of modes where $\mc{K} = \{1,2,1\}$. This means that the subset $\bs{s}_1$ will always have $k_1=1$ target modulated states within its pattern region, $\bs{s}_2$ will have $k_2 = 2$ and $\bs{s}_2$ will have $k_3 = 1$. An image space $\mc{U}_{\text{TPF}}^{\mc{S},\mc{K}}$ can then be generated according to these properties. The vast space of possible configurations puts Eve at a disadvantage if she cannot determine the precise modulation scheme. }}
\label{fig:LTPF_explain}
\end{figure}

{As an example, let us consider Fig.~\ref{fig:LTPF_explain}(a). This depicts an $m= 11$ mode coherent pattern space with a specific tripartite locality structure $\mc{S} = \{\bs{s}_1,\bs{s}_2,\bs{s}_3\}$ where $|\bs{s}_1| = |\bs{s}_3| = 3$ and $|\bs{s}_2| = 5$. We can attribute a $k$-TPF property to each of these subregions which will inform Bob how many target modulated modes he should expect within each subregion. If this information can be concealed from Eve, then secure rates can be enhanced by encoding information asymmetry. }

\subsubsection{Secure Rates}

\begin{figure}
\hspace{1cm}(a) \\
\hspace{5mm}\includegraphics[width=0.95\linewidth]{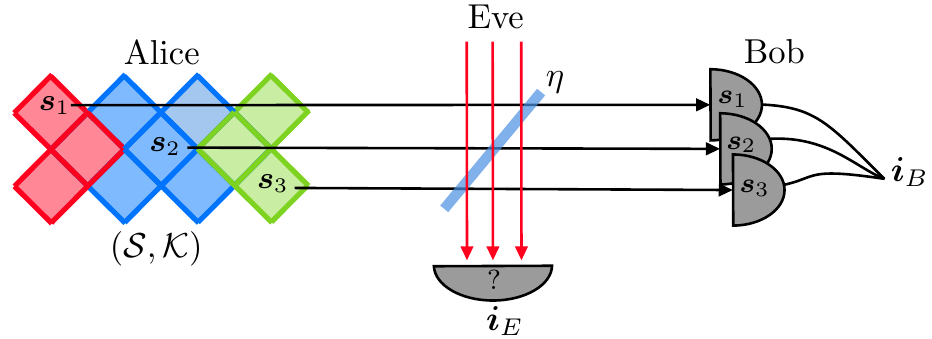}\\
\hspace{0.55cm} (b) \hspace{3.5cm} (c)\\
\includegraphics[width=\linewidth]{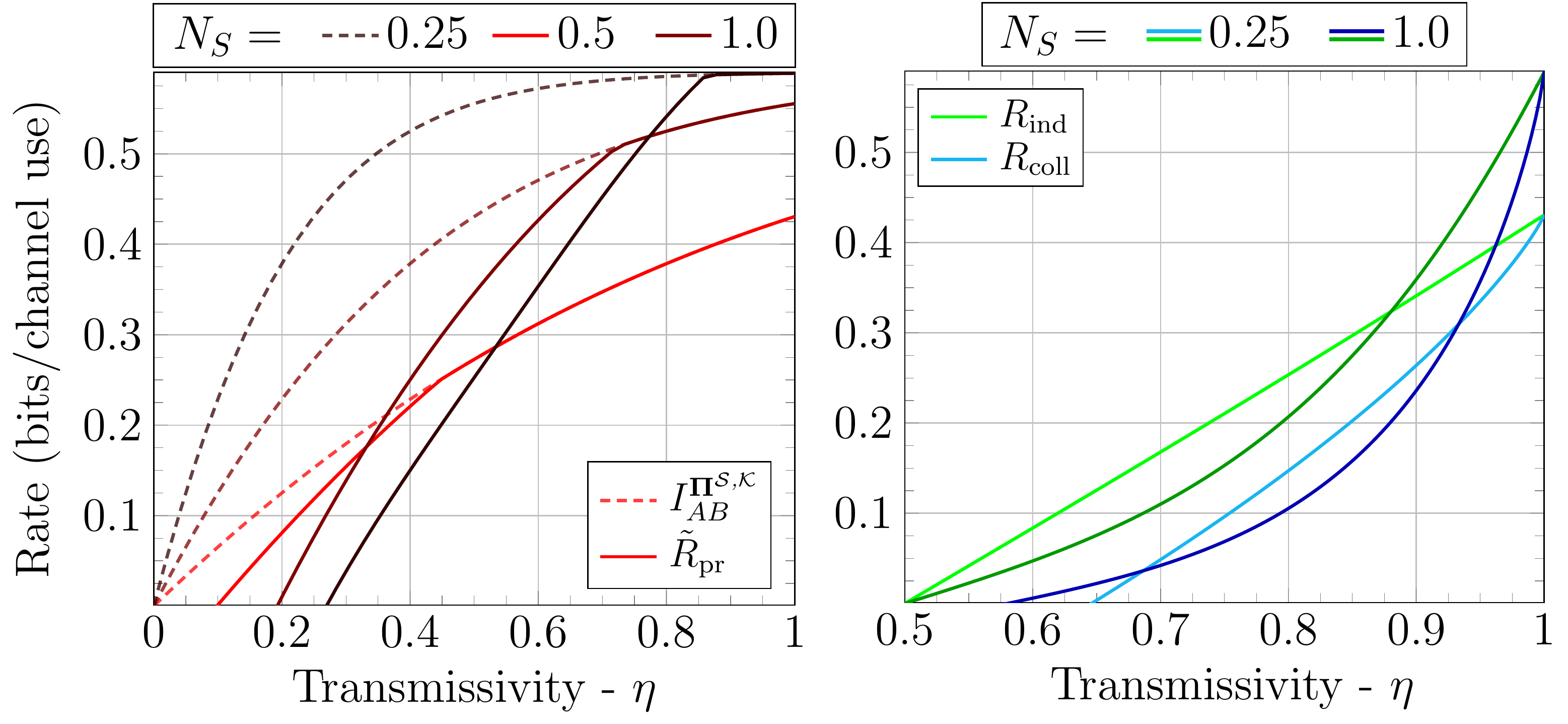}
\caption{Panel (a) illustrates LTPF Pattern Communication using the encoding scheme $(\mathcal{S},\mathcal{K}) =( \{\bs{s}_1,\bs{s}_2,\bs{s}_3\}, \{1,2,1\})$ described in Fig.~\ref{fig:LTPF_explain}. Panel (b) describes the behaviour with respect to transmissivity of the optimised mutual information between Alice and Bob (dashed), and the secure communication rate under probabilistic attacks $\tilde{R}_{\text{pr}}$ (solid). Panel (c) then describes secure communication rates from Eqs.~(\ref{eq:kTPF_Gen}) - (\ref{eq:kTPF_Ind}) considering non-approximate attacks, in which Eve's possesses information and resources that are as good or better than Alice and Bob.  }
\label{fig:LTPF_rates}
\end{figure}

Measurement outcome probabilities can be assessed for PGMs by means of Gram matrices. Here we define $G[\mathcal{U}]$ as the Gram matrix of an ensemble of lossy coherent pattern states that form the image space $\mathcal{U}$, 
\begin{equation}
G[\mathcal{U}]_{\bs{i},\bs{i}^{\prime}} = \braket{ \eta \alpha_{\bs{i}} | \eta \alpha_{\bs{i}^{\prime}}}, \text{ } \bs{i},\bs{i}^{\prime} \in \mathcal{U}.
\end{equation}
The square root of the Gram matrix of a pure state ensemble can be used to derive conditional probabilities of PGM measurement outcomes. For local sub-patterns of transmission states,
\begin{gather}
p(\bs{i}_B^{\bs{s}_j} | \bs{i}_A^{\bs{s}_j} ) = \left[\Big(\sqrt{G\big[\kCPF{|\bs{s}_j|,k_j}\big]}\Big)_{ \bs{i}_A^{\bs{s}_j}  \bs{i}_B^{\bs{s}_j}}\right]^2.
\end{gather}
Using Bayes theorem to find the converse conditional probabilities $p(\bs{i}_A^{\bs{s}_j} | \bs{i}_B^{\bs{s}_j})$, the conditional probability of Alice having transmitted a global pattern $\bs{i}_A$ given Bob reconstructed the global pattern $\bs{i}_B$ is given by, 
\begin{gather}
p(\bs{i}_A | \bs{i}_B)  = p \Big( \bigcap_{j=1}^n \bs{i}_A^{\bs{s}_j} \Big|  \bigcap_{j=1}^n \bs{i}_B^{\bs{s}_j}\Big) = \prod_{j=1}^n p(\bs{i}_A^{\bs{s}_j} | \bs{i}_B^{\bs{s}_j}) .
\end{gather}
The mutual information can then be computed as,
\begin{gather}
I_{AB}^{\bs{\Pi}^{\mathcal{S},\mathcal{K}} } \defeq \log \Sigma + \hspace{-4mm}\sum_{\bs{i}_A, \bs{i}_B \in \kCPF{\mathcal{S},\mathcal{K}}} \hspace{-0mm} \hspace{-2mm} p(\bs{i}_A , \bs{i}_B) \log\left( p(\bs{i}_A | \bs{i}_B)\right).
\end{gather}
This allows us to write the secure communication rates from Eqs.~(\ref{eq:GenRate}) and (\ref{eq:IndRate}), under collective and individual attacks respectively,
\begin{align}
R_{\text{coll}}  &= I_{AB}^{\bs{\Pi}^{\mathcal{S},\mathcal{K}} }(\eta)- \chi_{AE}(1-\eta), \label{eq:kTPF_Gen}\\
R_{\text{ind}} &= I_{AB}^{\bs{\Pi}^{\mathcal{S},\mathcal{K}} }(\eta)-  I_{AE}^{\bs{\Pi}^{\mathcal{S},\mathcal{K}} }(1-\eta).\label{eq:kTPF_Ind}
\end{align}
These rates assume Eve has full knowledge of the encoding scheme $(\mathcal{S},\mathcal{K})$, and can be seen in Fig.~\ref{fig:LTPF_rates}(c) for the specific encoding $(\mathcal{S},\mathcal{K}) =( \{\bs{s}_1,\bs{s}_2,\bs{s}_3\}, \{1,2,1\})$. These rates are of course secure for $\eta \gtrsim 0.5$. 

When considering a large number of modes $m$, there is a super-exponentially increasing number of ways in which $\mathcal{S}$ and $\mathcal{K}$ can be chosen (see Appendix \ref{sec:LTPFDegen}). Therefore, it is non-trivial to consider a scenario that Eve is not in full possession of this codebook, due to its highly degenerate characteristics.

The most threatening approximate attack is probabilistic, and is a situation in which Eve has deduced $\mathcal{S}$ (the locality structure), but is unaware of $\mathcal{K}$ (TPF of each sub-pattern).
In this case, Eve must optimise her measurement apparatus in order to comply with $\mathcal{S}$ but without imposing any bias on $\mathcal{K}$. If she is biased, then she risks utilising an image space that is missing essential codewords from the real codebook. Therefore her best strategy is to utilise a larger potential image space that is compliant with $\mathcal{S}$; then Eve should assume that the number of target states that she measures in each sub-pattern is consistent with the real $\mathcal{K}$. That is, Eve must infer $\mathcal{K}$ directly from her measurements. Hence, Eve constructs an image space which is a concatenation of all $\mathcal{S}$-locality adhering patterns,
\begin{equation}
\kCPF{\mathcal{S}}  = \bigcup_{j=1}^n \Big( \kCPF{|\bs{s}_j|,1} \cup \ldots \cup \kCPF{|\bs{s}_j|,|\bs{s}_j|-1} \Big) .
\label{eq:EveGenImS}
\end{equation}
Eve's image space (and thus the coherent state ensemble) no longer satisfies GUS since the $k$-TPF properties of each pattern region are now variable. However, she may still use PGMs, as the requirement that the $k$-TPF property of each pattern region falls within $k_j \in \{1,\ldots, |\bs{s}_j|-1\}$ means that she can rule out \textit{some} invalid patterns, allowing her to outperform local measurements. Eve's measurement operators are thus
\begin{equation}
\bs{\Pi}^{\mathcal{S}} = \left\{ \Pi_{\bs{i}} \right\}_{\bs{i}\in\kCPF{\mathcal{S}}} , \>\> \Pi_{\bs{i}}^{\mathcal{S}} = \bigotimes_{j=1}^{n} \Pi_{\bs{i}^{\bs{s}_j}}^{|\bs{s}_j|,\{1,\ldots,|\bs{s}_j|-1\}}.  \label{eq:EveGenPi}
\end{equation}

To analyse Eve's maximum mutual information we can use Gram matrices in accordance with the sub-optimal image space from Eq.~(\ref{eq:EveGenImS}), such that
\begin{gather}
I_{AE}^{\bs{\Pi}^{\mathcal{S}}} \defeq \log \Sigma + \hspace{-6mm}\sum_{\bs{i}_A \in \kCPF{\mathcal{S}, \mathcal{K}} , \bs{i}_E \in \kCPF{\mathcal{S}}} \hspace{-6mm} p(\bs{i}_A , \bs{i}_E) \log\left( p(\bs{i}_A | \bs{i}_E)\right).
\end{gather}
Eve's unbiased strategy means that she may still discriminate patterns that do not exist within the correct codebook, leading to the inferior conditional entropy term above.

Furthermore, Eve will only obtain this information $I_{AE}^{\bs{\Pi}^{\mathcal{S}}}$ probabilistically, since it relies on her ability to correctly infer the $k$-TPF properties of the pattern space, $\mathcal{K}$. The probability of successful inference can also be computed via the Gram matrices of all the potential $k$-TPF sub-pattern ensembles, which we label $p_{\text{dec}}^{\mathcal{K}|\mathcal{S}}$ (see Appendix \ref{sec:kTPF_Inf}). Ultimately, her approximate attack results in the following secure communication rate,
\begin{align}
\tilde{R}_{\text{pr}}  &= I_{AB}^{\bs{\Pi}^{\mathcal{S},\mathcal{K}} }(\eta)-  p_{\text{dec}}^{\mathcal{K}|\mathcal{S}} \left( I_{AE}^{\bs{\Pi}^{\mathcal{S}} }(1-\eta)\right). \label{eq:kTPF_PrRate}
\end{align}
Note Eve's non-biased approach is much more effective than any guessing type scheme, since the number of ways in which Eve could choose $\mathcal{K}$ for large $m$ would quickly force $p_{\text{dec}}^{\mathcal{K}|\mathcal{S}} \rightarrow 0$.\par
Results for $\tilde{R}_{\text{pr}}$ are shown in Fig.~\ref{fig:LTPF_rates}(b). The undesirable contribution of invalid pattern states in $\kCPF{\mathcal{S}}$ clearly degrade Eve's information retrieval, resulting in a secure rate over much larger transmissivity intervals. These secure regions may be as low as $\eta \sim 0.1$ for signal energies $N_S = 0.25$. As the mean photon number $N_S$ is increased, Eve's inference abilities improve, causing the protocol to once more become less secure at lower transmissivities.

\subsection{Degenerate Encoding and Pattern Recognition}
The previous pattern modulation scheme example utilised a one-to-one encoding, attempting to exploit information asymmetry between Bob and Eve in order to obtain superior discriminatory measurements. In the following, we take a data-driven approach in which information is packaged through classifiable, degenerate patterns. It is then meaningful to consider a diminished approximate attack, such that overwhelming amounts of data have forced Eve into a limited resource position.\par

\subsubsection{Pattern Modulation Scheme}
As an example, we use the MNIST data-set to construct a degenerate pattern encoding method. This contains a data-set of $m = 28\!\times\!28$ pixel images $\bs{i}$, which can be classified as a decimal handwritten digit, formulating a 10-symbol alphabet $\mathcal{A} = \{0,\ldots,9\}$. The typical data-set is grey-scale, but the images can be polarised so to represent the modulation of a binary coherent state basis. Here, we utilise the MNIST training set $\mathcal{T} = \{ c_j ; \bs{i}_j \}_{j}$ to formulate our codebook, which contains an image space of $|\mathcal{T}| = 60000$ patterns, each of which have been pre-labelled with an exact classifier, $c$. Clearly, $|\mathcal{T}| \gg \mathcal{A}$, leading to a vastly degenerate codebook,
\begin{equation}
\mathcal{C} = \left\{ \big(c(\bs{i})~ ; \ket{\alpha_{\bs{i}}}\big) ~\big|~ (c(\bs{i}); \bs{i}) \in \mathcal{T} \right\}. \label{eq:MNISTCodebook}
\end{equation}

The modulation scheme proceeds as follows: Symbols can be encoded by ``drawing" a handwritten digit using binary-modulated coherent states. Alice can randomly generate and transmit these pattern to Bob through multi-mode pure-loss channels, who then uses a set of measurements $\bs{\Pi}$ to generate a noisy reconstruction of the pattern. Bob can consult with his codebook $\mathcal{C}$, and a (possibly pre-trained) classifier $\tilde{c}_{\mathcal{T}}$ (whose efficiency is dependent on the quality of $\mathcal{T}$) in order to decode the pattern. Simultaneously, we may consider an eavesdropper who applies a global beam-splitter attack to steal information, and may offer a variety of security threats based on her resources.

\begin{figure}[t]
\hspace{-3mm} (a) \\
\includegraphics[width=0.95\linewidth]{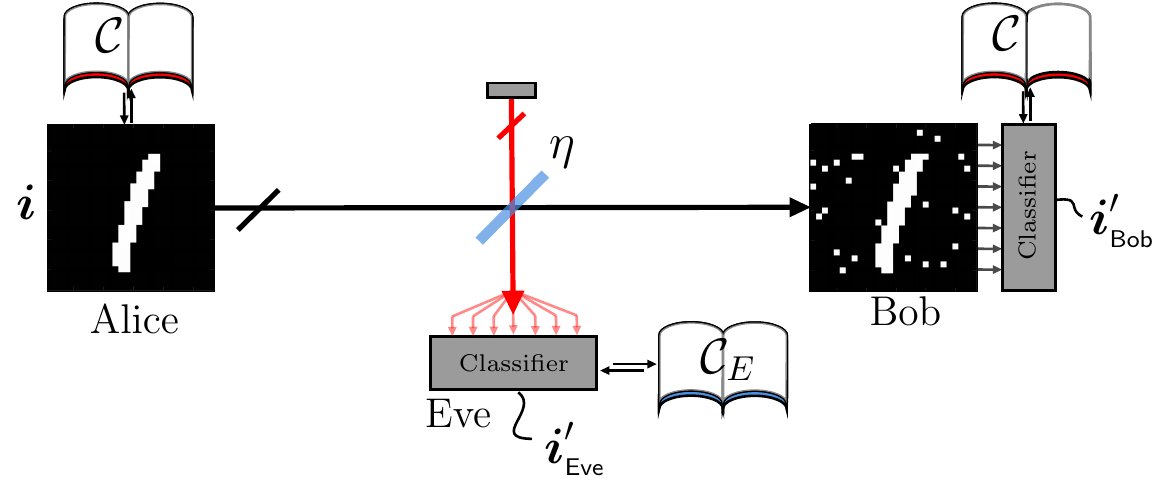}\\
\hspace{-3mm} (b) \\
\includegraphics[width=0.95\linewidth]{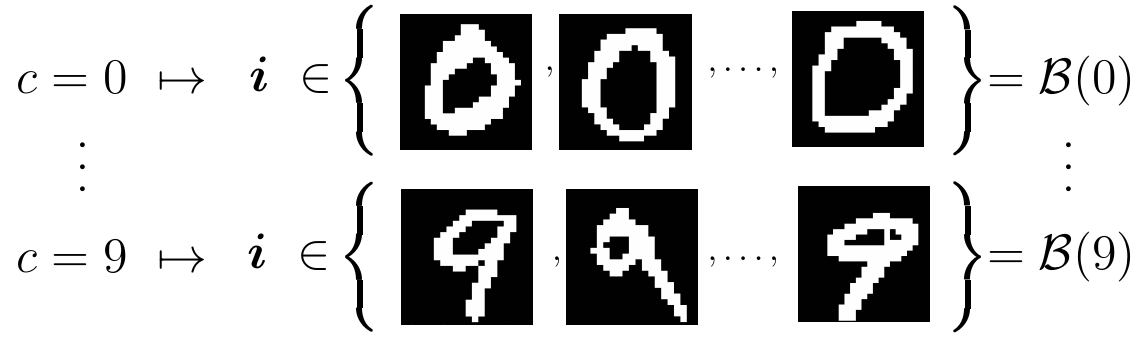}
\caption{Panel (a) describes pattern communication using a degenerate codebook. Alice and Bob possess some degenerate codebook $\mathcal{C}_{AB}$, while Eve possesses a potentially inferior codebook $\mathcal{C}_E \subseteq \mathcal{C}_{AB}$. Local measurements are employed in conjunction with a classifier. Panel (b) displays an example of degenerate coding using the MNIST handwritten digit dataset, with $d \in \{0,\ldots,9\}$ symbols encoded into coherent patterns that explicity ``draw" these digits. }
\end{figure}

The large number of modes $m = 784$, and the non-uniformity of MNIST patterns makes it very difficult to determine optimal measurements. This of course motivates the use of local receivers assisted by statistical classifiers.
Hence, we assume that Bob performs local Helstrom measurements (e.g.~via a Dolinar receiver \cite{Dolinar}), denoting the associated POVM as $\bs{\Pi}^{\otimes} \defeq \bigotimes_{j=1}^{m} \Pi_{i_j}$. Noisy patterns can be simulated by performing single pixel bit flips on each mode in a transmitted pattern with probability,
\begin{equation}
p_{\text{err}}^{\text{mode}} = \frac{1-\sqrt{1-e^{-\kappa\eta N_S}}}{2},
\end{equation}
where $\kappa = 4$ ($\kappa = 1$) for BPSK (BAM).\par
\begin{figure}
\hspace{8mm} (a) MI and Symmetric Attacks\\
\includegraphics[width=0.8\linewidth]{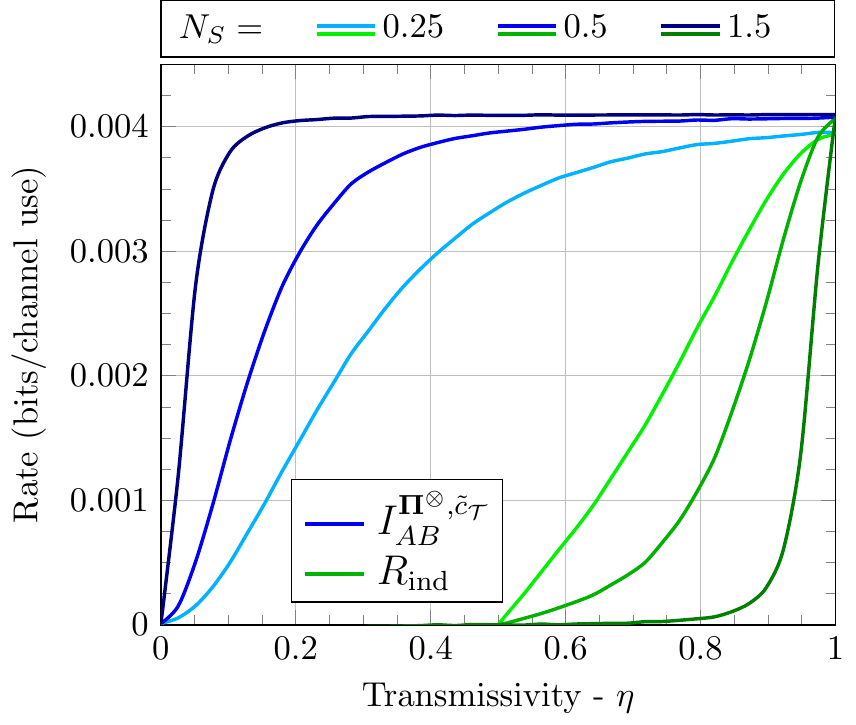}\\
\hspace{8mm} (b) Minimum Approximate Attacks\\
\includegraphics[width=0.8\linewidth]{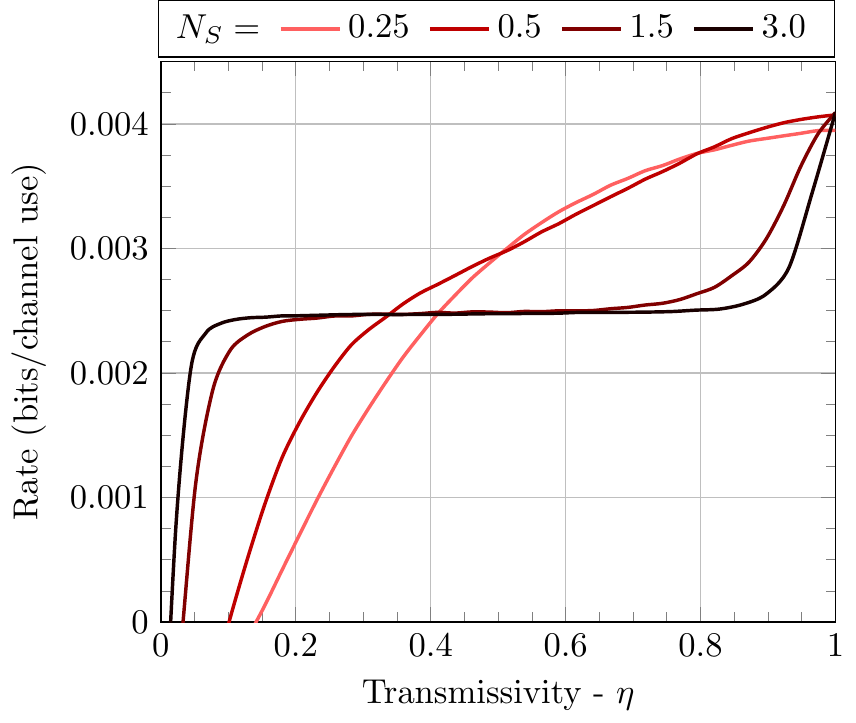}
\caption{MNIST Degenerate pattern communication. Panel (a) depicts the simulated mutual information (blue) and symmetric rate from Eq.~(\ref{eq:MNIST_IndRate}) (green), while Panel (b) computes the rate under minimum approximate attacks given in Eq.~(\ref{eq:MNIST_dim}). Both Bob and Eve employ the use of CNN decoders, and we consider communication of patterns with $N_S \in \{0.25,0.5,1.5,3.0\}$, with simulated communication rates computed over $|\mathcal{V}|=10000$ transmissions, and averaged over 20  simulations.}
\label{fig:MNIST_rates}
\end{figure}

There are a plethora of potential classifiers that can be used in this communication setting, ranging from simple nearest neighbour classifiers, to more sophisticated Convolutional Neural Networks (CNN). In this work we utilise shallow CNNs which act as neural decoders. CNNs are a very popular tool for image processing and pattern recognition, due to their high performance classification accuracies even amidst noisy inputs, and therefore pose as an excellent model classifier for Bob and/or Eve \cite{nielsenNN}.

\subsubsection{Secure Rates}

The MNIST data-set also contains an evaluation set $\mathcal{V} = \{c_k; \bs{i}_k\}_k$ with $|\mathcal{V}| = 10000$ patterns and their precise classification. Importantly, these are completely independent samples from the training set, $\mathcal{V} \cap \mathcal{T} = \varnothing$, and can therefore be used to empirically simulate and evaluate communication over $|\mathcal{V}|$ transmissions.

Let $c_A$ and $c_B$ denote the class of Alice's transmission and the class inferred by Bob's classification procedure respectively, $c_A,c_B \in \{0,\ldots,9\}$. The conditional probability of having transmitted a message with classification $c_A$, given Bob has used a classifier $\tilde{c}$ to infer $c_B$ can be approximated using $\mathcal{V}$, 
\begin{equation}
p(c_A | c_B) = \frac{p(c_A,c_B)}{p(c_B)} \approx  \frac{\sum_{(c(\bs{i});\bs{i})\in\mathcal{V}} \delta(c(\bs{i}), c_A)  \delta(\tilde{c}_{\mathcal{T}}(\bs{i}), c_B)}{\sum_{(c(\bs{i});\bs{i})\in\mathcal{V}}  \delta(\tilde{c}_{\mathcal{T}}(\bs{i}), c_B)},
\end{equation}
where $\delta$ is a Kronecker delta function $\delta(c_j, c_k) = 1$ iff the classifications $c_j = c_k$. Using these approximate probability distributions we may compute the mutual information between Alice and Bob,
\begin{align}
I_{AB}^{\bs{\Pi}^{\otimes},\tilde{c}_{\mathcal{T}}} \approx \log 10 -  \hspace{-2.5mm}\sum_{c_A, c_B \in\mathcal{A}} p(c_A,c_B) \log\left( {p(c_A|c_B)}\right).
\end{align}
This approximates their average mutual information over $|\mathcal{V}|$ transmissions, and can be seen in Fig.~\ref{fig:MNIST_rates}(a). \par
The role of an eavesdropper can now be investigated. Once again, in a worst case scenario Eve may capture and store her share of all incident modes in a quantum memory, and extract the accessible information via an optimal collective attack. For such a large, degenerate ensemble of quantum states this is an expensive, and potentially unrealistic tactic (certainly for near-term technologies). Furthermore, computing the Holevo information in this context is extremely demanding for the same reasons, and thus we leave this security consideration to future studies \footnote{Nonetheless, it would be interesting to investigate security conditions granted Eve utilises \textit{imperfect} quantum memories and realistic data-processing times.}.\par
Alternatively we may consider the impact of individual attacks. In an informationally symmetric setting, Eve is aware of the codeword to alphabet mapping, and possesses an identical codebook $\mathcal{C}_E = \mathcal{C}$. The secure communication rate will thus follow Eq.~(\ref{eq:IndRate}), 
\begin{equation}
R_{\text{ind}} = I_{AB}^{\bs{\Pi}^{\otimes},\tilde{c}_{\mathcal{T}}}(\eta) - I_{AE}^{\bs{\Pi}^{\otimes},\tilde{c}_{\mathcal{T}}}(1-\eta). \label{eq:MNIST_IndRate}
\end{equation}
This symmetric rate is shown in Fig.~\ref{fig:MNIST_rates}(a), which follows the typical behaviour for communication in direct reconciliation, and only admits security for $\eta \geq 0.5$.   \par
However, for a codebook of this magnitude, it is not trivial to assume that an arbitrary eavesdropper can obtain perfect encoding knowledge. Indeed, it is non-trivial to consider scenarios such that (\textit{i}) Eve does not possess the same resources as Bob $\mathcal{T}_E \neq \mathcal{T}$, or (\textit{ii}) Eve does not possess the codebook at all. We may simulate rates based on the assumption in (\textit{i}), and outline a generic adaptive protocol for Eve's worst case scenario in (\textit{ii}). \par
The assumption that Eve possesses the codebook mapping, but only partial resources leads to a diminished approximate attack, where Eve's training set may now be considered as a subset $\mathcal{T}_E \subset \mathcal{T}$. This separation in training set quality will render Eve's classifier $\tilde{c}_{\mathcal{T}_E}$ inferior with respect to Bob's $\tilde{c}_{\mathcal{T}}$ especially when $|\mathcal{T}| \gg |\mathcal{T}_E|$. This results in a rate described by
\begin{equation}
\tilde{R}_{\text{dim}} = I_{AB}^{\bs{\Pi}^{\otimes},\tilde{c}_{\mathcal{T}}}(\eta) - I_{AE}^{\bs{\Pi}^{\otimes},\tilde{c}_{\mathcal{T}_E}}(1-\eta). \label{eq:MNIST_dim}
\end{equation}
For an eavesdropper who is solely aware of the codeword mapping, they will only possess single examples of each codeword such that $|\mathcal{B}(c)| = 1$, $\forall c$, and their training set $|\mathcal{T}_E| =10$. This defines a \textit{minimum approximate attack}, since this is the minimum amount of information Eve needs to apply a deterministic attack. Results for this rate are shown in Fig.~\ref{fig:MNIST_rates}(b). As expected, Eve's restricted resources lead to a dramatically more secure protocol, allowing Alice and Bob to communicate securely at much lower transmissivities. As the mean photon energy $N_S$ is increased, the rate begins to plateau with respect to transmissivity; improvements in Eve's single mode discrimination is incapable of boosting her classification performance until $\eta\sim 0$. This lets Alice and Bob achieve a near constant non-zero rate within a large window of transmissivities.\par
Finally, one can consider the strategy of a completely ignorant eavesdropper. Now Eve knows nothing about the encoding, and must construct her own codebook in order to extract any information at all. To do so, Eve must observe transmissions from the evaluation set and try to infer an approximate alphabet $\tilde{\mathcal{A}}$ and its respective codeword mappings. This can be achieved (albeit with some difficulty when transmissions are particularly noisy) by means of a data-clustering algorithm over the span of many transmissions, and can then be used to devise an approximate codebook and classifier. This will result in a probabilistic form of Eq.~(\ref{eq:MNIST_dim}) with a decoding error associated with alphabet inference. In the limit of many transmissions, this strategy may have some success, but will still result in a very secure rate for Alice and Bob.

\section{Discussion \label{sec:Concl}}
We have investigated a multi-mode modulation scheme for bosonic quantum communications. We have shown that is possible to encode information into multi-mode coherent states which are discretely-modulated according to specific structures, which we name quantum patterns. Likening the task of communication with pattern recognition, we study abstract encodings based on collections of coherent quantum patterns which may possess extreme degeneracies and non-linearities. From this, interesting questions regarding practical/realistic security emerge. We elucidate these general arguments with some example pattern encodings: One of which exploits eavesdropper ignorance to obtain superior quantum measurements, while the other employs degenerate coding in order to capitalise on an eavesdropper's limited resources.\par

These methods and results are informative to the fact that multi-mode encoded information can be used to introduce serious complications for eavesdroppers. In particular, the versatility of trainable classifiers in cooperation with arbitrarily complex (even adaptive) coding schemes could be used to introduce novel layers of security in quantum communication protocols. \par

There are clearly many immediate possible developments such as the explicit investigation of $k$-ary modulated patterns, and the extension to reverse reconciliation protocols. It would also be valuable to better understand the abilities of an eavesdropper when exposed to a large, degenerate code. If an attacker's resources for pattern inference can be securely limited, then their threat can be minimised, even when in possession of a quantum memory. This would require an upper bound on Eve's classification power via generally quantum resources, given she has extracted the accessible information. Analyses from \cite{QMLGen,PreskillQML} may be of use for this.\par

In this work, we have focussed on the use of quantum patterns constructed from coherent states. This was carried out as an expedient translation from the most common and practical CV-QKD protocols. Furthermore, coherent state discrimination and its error rates are well understood. Yet in general, quantum pattern states can be constructed using any kind of locally modulated states, such as thermal states, squeezed states etc. To this end, it would be interesting to explore the incorporation of \textit{entangled} quantum pattern states, which would exploit non-local modulations to construct global patterns. 
Entanglement-assistance is well known to be a powerful resource for quantum communications \cite{Bennett92,Holevo02,ZhuangAdditive17,Guo_EntComms}, and in this setting might be possible to introduce further complications for eavesdroppers.

Most importantly, using pattern encoding in order to enhance secure protocols against collective attacks (rather than individual) poses the greatest reward. Devising a secure training protocol for the classifiers of trusted parties would allow for the benefits of approximate attacks to be realised within this stricter framework. The covert incorporation of information asymmetry between users and eavesdroppers in QKD could be of great benefit to security, posing a fascinating future investigative path.

\section*{Acknowledgments}
C.H acknowledges funding from the EPSRC via a Doctoral Training Partnership (EP/R513386/1). S.P acknowledges funding from the European Union’s Horizon 2020 Research and Innovation Action under grant agreement No.~862644 (Quantum readout techniques and technologies, QUARTET).


%

\appendix

\section{Discrimination via PGMs \label{sec:kTPF_Inf}}
Consider an ensemble of coherent pattern states through a uniform lossy multi-channel $\{p_{\bs{i}} ; \alpha_{\bs{i}}^{\eta}\}_{\bs{i}\in\mc{U}}$ where $\mc{U}$ is an image space, and assuming equal \textit{a priori} probabilities for each pattern to occur $p_{\bs{i}} = 1/|\mc{U}|$ for all $\bs{i} \in \mc{U}$. Since the ensemble is constituent of pure states $\ket{\alpha_{\bs{i}}^{\eta}}$ we can use its Gram matrix in order to study the effectiveness of discrimination using Pretty Good Measurements (PGMs), whose elements takes the form
\begin{equation}
G[\mathcal{U}]_{\bs{i},\bs{i}^{\prime}} = \braket{ \eta \alpha_{\bs{i}} | \eta \alpha_{\bs{i}^{\prime}}},
\end{equation}
The average error probability of discrimination is then given by
\begin{equation}
p_{\text{err}}[\mathcal{U}] \defeq 1 - \frac{1}{|\mathcal{U}|} \left(\sum_{i=1}^{|\mathcal{U}|} \lambda^{\frac{1}{2}}_i\right)^2 \label{eq:perrGUS} ,
\end{equation}
where $\{ \lambda_i \}_{i=1}^{|\mc{U}|}$ are the eigenvalues of the Gram Matrix.
This represents the average error probability of discriminating any pattern $\bs{i}\in\mc{U}$ from all the other patterns in this image space. 

\section{Decoding LTPF Modulation Schemes}
{
In this Appendix we derive important quantities used in the study of LTPF encoding schemes. In particular, we derive Eves decoding probability of the $k$-TPF partition set $\mc{K}$ given that she has knowledge of $\mc{S}$, used in the main text to compute Alice and Bob's secure rate under probabilistic attack. Furthermore, we discuss the degeneracy of LTPF encodings, which preclude (or make difficult) effective inference methods of $\mc{S}$ or $\mc{K}$.
}
\subsection{Probability of Inferring $\mc{K}$ given $\mc{S}$}

{
Consider LTPF pattern communication as in the main text, where the modulation scheme is completely characterised by a locality partition set $\mc{S}$ (which describes how Alice and Bob choose regions within the pattern states to encode information) and a $k$-TPF partition set $\mc{K}$ (which describes how many target modes and background modes will be present within any given sub-region of the pattern states). If Eve has knowledge of $\mc{S}$ but not $\mc{K}$, then her information retrieval is disadvantaged as she cannot fully optimise her discriminatory measurements. But worse than this, Eve must deduce the properties of the $\mathcal{K}$ if she is to steal any information at all, as it is required to properly decode any encoded classical information from her collected states. 
}

{Let us consider Eve's scenario. Alice generates $m$-mode coherent patterns states according to the modulation scheme $(\mc{S},\mc{K})$, which Eve intercepts. Since Eve has knowledge of $\mc{S}$ she knows that she should apply $|\bs{s}_j|$-mode PGMs over each sub-pattern of the global state she recovers from the beamsplitter. The problem is that she is unable to fully optimise these measurements because she does not know the precise number of target modulated modes $k_j$ within each sub-pattern state. 
As discussed in Section \ref{sec:LTPF}, we know that valid target numbers in a pattern region $\bs{s}_j$ belong to the set of values $\bs{k} = \{1,\ldots,|\bs{s}|_j-1\}$ to ensure at least a binary variable is encoded in each sub-pattern. Hence, Eve must utilise measurements that account for a variable amount of target modes at each sub-pattern. This non-biased approach means that she must consider her output ensemble to be generated by the image space,
\begin{equation}
\kCPF{\mathcal{S}}  = \bigcup_{j=1}^n \bigcup_{k=1}^{|\bs{s}_j|-1} \kCPF{|\bs{s}_j|,k},
\end{equation}
as in Eq.~(\ref{eq:EveGenImS}) in the main text. This image space contains $\Sigma(\bs{s}_j) \defeq  {\sum_{i=1}^{|\bs{s}_j|-1} C_{|\bs{s}|_j}^{i}}$ potential output states at each sub-pattern. }

{Eve performs these measurement to discriminate the pattern states. However she must further infer $k_j$ over each sub-region $\bs{s}_j$ in order to decode the transmissions into their binary representations. In the absence of prior knowledge of $\mc{K}$, we consider Eve's strategy to be direct inference of $k_j$ from her discrimination. That is, if Alice transmits a sub-pattern $\bs{i}_A^{\bs{s}_j}$ with $k_j$ target modulated modes which Eve discriminates as $\bs{i}_E^{\bs{s}_j}$ with $\tilde{k}_j$ target modulations, she must infer that $\tilde{k}_j$ is the correct value in the encoding scheme. This allows Eve to build up an approximate $k$-TPF partition set $\tilde{\mc{K}} = \{\tilde{k}_1, \ldots,\tilde{k}_N\}$ associated with each transmission. }

{The question is thus: What is the probability that Eve correctly infers $\tilde{\mc{K}} = \mc{K}$? This is equivalent to asking: What is the probability that Eve discriminates her intercepted pattern state as belonging to the correct image space $\mc{U}_{\text{TPF}}^{\mc{S},\mc{K}}$. Consider a single sub-pattern $\bs{s}_j$ with a true number of target modulations $k_j$. The average error probability of Eve inferring a target modulation $\tilde{k}_j$  is equal to}
\begin{equation}
{p}(\tilde{k}_j|k_j,{\bs{s}_j}) = \sum_{\bs{i}_A^{\bs{s}_j}\in\kCPF{|\bs{s}_j|,k_j}}  \sum_{\bs{i}_E^{\bs{s}_j} \in\kCPF{|\bs{s}_j|,\tilde{k}_j}} \frac{p(\bs{i}_E^{\bs{s}_j} |\bs{i}_A^{\bs{s}_j})}{|\kCPF{|\bs{s}_j|,k_j}|}  
\end{equation}
{Since we are using PGMs and coherent pattern states transmitted through pure-loss channels, we can replace the conditional probability $p(\bs{i}_E^{\bs{s}_j} |\bs{i}_A^{\bs{s}_j})$ with its computable value}
\begin{align}
p(\bs{i}_E^{\bs{s}_j} |\bs{i}_A^{\bs{s}_j}) &= \text{Tr}\left[ \Pi_{\bs{i}_E^{\bs{s}_j}} \alpha_{\bs{i}_A^{\bs{s}_j}}^{1-\eta} \right],\\
&= \left[\Big(\sqrt{G\big[\kCPF{\mc{S}}\big]}\Big)_{ \bs{i}_A^{\bs{s}_j}  \bs{i}_E^{\bs{s}_j}}\right]^2.
\end{align}

{Therefore, the average success probability of Eve inferring $k_j$ can be computed by summing the error probabilities $p(\tilde{k}_j|k_j, \bs{s}_j)$ over all the values that she believes $\tilde{k}_j$ could possibly take. More precisely,}
\begin{equation}
{p}_{\text{dec}}^{k_j|\bs{s}_j} \defeq 1 - \sum_{\tilde{k}_j = 1}^{|\bs{s}_j|-1} {p}(\tilde{k}_j|k_j,{\bs{s}_j})
\end{equation}
{Since she has to do this for all sub-patterns, we can then finally compute the successful decoding probability of inferring $\mc{K}$ directly from her measurements,}
\begin{equation}
p_{\text{dec}}^{\mc{K}|\mc{S}} \defeq \prod_{j=1}^n {p}_{\text{dec}}^{k_j|\bs{s}_j}.
\end{equation}

{Hence, $p_{\text{dec}}^{\mathcal{K}|\mathcal{S}}$ is the success probability of inferring $\mathcal{K}$ through PGM measurements which are non-biased to the number of target modes in each sub-region $\bs{s}_j$, given that the locality structure $\mathcal{S}$ is already known. There may exist more sophisticated methods that Eve can employ to more accurately infer the partition set $\mc{K}$. Nonetheless, this offers an insightful inspection into the effects that information asymmetry has on communicators and attackers.}

\subsection{Degeneracy of LTPF Encoding Schemes \label{sec:LTPFDegen}}
Here we briefly summarise the degeneracy properties of LTPF encoding schemes, and the number of ways that a specific $(\mathcal{S},\mathcal{K})$ pair can be chosen over $m$-mode patterns. A partition of $m \in \mathbb{N}$ into $n$ parts is defined as an ordered vector $\bs{x} = \{x_1,\ldots,x_n\}$ where $x_j \in \mathbb{N}$, $x_1 \geq \ldots \geq x_n > 0$ and $\sum_{j=1}^n x_j = m$. We denote this as $\bs{x} \vdash_n m$. Given the multinomial coefficient
\begin{equation}
M_m^{\bs{x}} = M_{m}^{x_1\ldots,x_n} \defeq \prod_{j=1}^n C_{m - \sum_{k=1}^j x_k}^{x_j},
\end{equation} 
we define a modification which discards permutations that are invariant under the shuffling of sub-patterns \footnote{For example, it is clear that the locality structures $\{\{1,2\},\{3,4\},\{5,6,7,8\}\}$ and $\{\{3,4\},\{1,2\},\{5,6,7,8\}\}$ are the same, as subsets have just been relabelled.},
\begin{equation}
\tilde{M}_m^{\bs{x}} = \frac{M_m^{\bs{x}} }{\prod_{l=1}^{\max(\bs{x})}\left[ \sum_{k=1}^r \delta(x_k,l)\right]!} .
\end{equation}
where $\delta(x,y)$ is an integer Kronecker delta function.\par 
The number of ways that one may choose a locality partition set $\mathcal{S}$ over $m$-modes may be calculated using the above formalism, summing over all possible combinations and partitions. A simpler computation is given by the associated Stirling numbers of the second kind, which count the number of ways to partition $m$-modes into $n$ parts with minimum subset size $k$. These numbers obey the recurrence relation,
\begin{align}
S_{k}^{m}(n) = n S_{k}^{m-1}(n)+{C_{m-1}^{k-1}} S_{k}^{m-k}(n-1).
\end{align}
Restricting sub-pattern dimensions to $2 \leq |\bs{s}_j| \leq m$ (to ensure all sub-patterns can encode at least one bit), then the degeneracy of $\mathcal{S}$ is 
\begin{equation}
\mathcal{G}_{\mathcal{S}} \defeq \sum_{n=1}^{\lfloor m/2\rfloor} S_{2}^m(n) =  \sum_{n=1}^{\lfloor{{m}/{2}}\rfloor} \sum_{\bs{x}\vdash_n m} \hspace{-1mm}\tilde{M}_{m}^{\bs{x}}. \label{eq:Gs}
\end{equation}\par
The parallel freedom of locality and TPF-partition sets expands the space of encodings even further. For each sub-pattern $\bs{s}_j \in \mathcal{S}$, there will exist $\prod_{j=1}^n( |\bs{s}_j|-1)$ choices of $k_j$ target modulations, with the constraints of $2 \leq |\bs{s}_j| \leq m$ and $k \in \{1,\ldots,|\bs{s}_j|-1\}$. The total degeneracy of all possible schemes is then given by,
\begin{equation}
\mathcal{G}_{\mathcal{S},\mathcal{K}} \defeq \sum_{n=1}^{\lfloor{{m}/{2}}\rfloor} \sum_{\bs{x}\vdash_n m} \hspace{-1mm}\tilde{M}_{m}^{\bs{x}} \prod_{j=1}^n (x_j-1). \label{eq:Gsk}
\end{equation}\par
It is also useful to determine conditional degeneracies, $\mathcal{G}_{\mathcal{S}|\mathcal{K}}$ and $\mathcal{G}_{\mathcal{K}|\mathcal{S}}$ based on some leaked information that Eve may have obtained. If Eve is aware of $\mathcal{K} = \{k_1,k_2,\ldots,k_n\}$ only, she can still glean some information about $\mathcal{S}$. Thanks to $\mathcal{K}$, Eve can infer the number of sub-patterns $n$ and also the minimum size of the $j^{\text{th}}$ sub-pattern, $x_{\text{min}}(j) = k_j+1$. 
Alternatively, if Eve is only aware of $\mathcal{S}$, then she can significantly narrow the space of possible $\mathcal{K}$. She knows it consists of $n$-elements, and is aware of the maximum/minimum target numbers of each sub-pattern. We can then summarise the conditional degeneracies,
\begin{equation}
\mathcal{G}_{\mathcal{S}|\mathcal{K}} \defeq  \hspace{-3mm}\sum_{\substack{\bs{x}\vdash_n m ,\\ x_j \geq k_j+1, \forall j}} \hspace{-3mm}\tilde{M}_{m}^{\bs{x}} , ~~~\mathcal{G}_{\mathcal{K}|\mathcal{S}} \defeq \prod_{j=1}^n( |\bs{s}_j|-1).
\end{equation}

In general, $\mathcal{G}_{\mathcal{S}|\mathcal{K}} \gg \mathcal{G}_{\mathcal{K}|\mathcal{S}}$, hence it is always more secure to keep $\mathcal{S}$ secret. Regardless, one can always choose a locality structure  $\mathcal{S}$ which maximises this degeneracy. Interestingly, one finds that constraining the number of sub-patterns sizes as $|\bs{s}_j| \in \{4, 5\}$, and maximising the number of sub-patterns with $|\bs{s}_j|=5$, produces the desired result. Defining the following function,
\begin{equation}
g_{\mathcal{K}|\mathcal{S}}(m) \defeq
\begin{cases}
m -1 &\text{ if }  2 \leq m \leq 7, \\
4^{\frac{m}{5}} &\text{ if } (\frac{m}{5} \in \mathbb{N}) \land (m > 7),\\
4^{\lceil\frac{m}{5}\rceil}\left(\frac{3}{4}\right)^{5-(m\>\text{mod}\>5)} &\text{ otherwise}.
\end{cases}
\end{equation}
we can write,
\begin{equation}
\mathcal{G}_{\mathcal{K}|\mathcal{S}} \leq \max_{\mc{S}} \mc{G}_{\mc{K}|\mc{S}} = g_{\mathcal{K}|\mathcal{S}}(m).
\end{equation}

\end{document}